\documentclass[traditabstract,twocolumn]{aa} % for the letters 
\usepackage{txfonts,amssymb,wasysym,textcomp,graphicx,natbib,twoopt,lscape}
\usepackage[breaklinks=true]{hyperref} %% to avoid \citeads line fills 
\bibpunct{(}{)}{;}{a}{}{,} %% natbib format like A&A and ApJ 
\newcommandtwoopt{\citeads}[3][][]{\href{http://adsabs.harvard.edu/abs/#3}% 
{\citealp[#1][#2]{#3}}} 
\newcommandtwoopt{\citepads}[3][][]{\href{http://adsabs.harvard.edu/abs/#3}% 
{\citep[#1][#2]{#3}}} 
\newcommandtwoopt{\citetads}[3][][]{\href{http://adsabs.harvard.edu/abs/#3}% 
{\citet[#1][#2]{#3}}} 
\newcommandtwoopt{\citeyearads}[3][][]% 
{\href{http://adsabs.harvard.edu/abs/#3}{\citeyear[#1][#2]{#3}}}
\usepackage{graphicx}
\usepackage{longtable}
\usepackage{natbib}
\bibpunct{(}{)}{;}{a}{}{,} % to follow the A&A style
\usepackage{txfonts}

\newcommand{\rxte}{{\it RXTE}}

\newcommand{\suzaku}{{\it Suzaku}}
\newcommand{\xmm}{{\it XMM}-Newton}

\newcommand{\swift}{{\it Swift}}

\newcommand{\nustar}{\textit{NuSTAR}}

\newcommand{\ms}{$M_{\odot}$}

\newcommand{\lumcgs}{ergs~s$^{-1}$}
\newcommand{\chisq}{$\chi^{2}_{\nu}$}
\newcommand{\nodata}{...}

\begin{document}

   \title{Dramatic change in the boundary layer in the symbiotic recurrent nova T Coronae Borealis.}

   \author{G. J. M. Luna, 
          \inst{1,2,3}, 
          K. Mukai,
           \inst{4,5}
           J. L. Sokoloski,
           \inst{6}
           T. Nelson, 
           \inst{7}
           P. Kuin,
           \inst{8}
           A. Segreto,
          \inst{9}
           G. Cusumano,          
          \inst{9}
          M. Jaque Arancibia,
          \inst{10,11}
           \and
         N. E. Nu\~nez,
           \inst{11}
         }

   \institute{CONICET-Universidad de Buenos Aires, Instituto de Astronom\'ia y F\'isica del Espacio, (IAFE), Av. Inte. G\"uiraldes 2620, C1428ZAA, Buenos Aires, Argentina\\
              \email{gjmluna@iafe.uba.ar}
                  \and
                  Universidad de Buenos Aires, Facultad de Ciencias Exactas y Naturales, Buenos Aires, Argentina
                  \and
                  Universidad Nacional Arturo Jauretche, Av. Calchaqu\'i 6200, F. Varela, Buenos Aires, Argentina
                  \and
CRESST and X-ray Astrophysics Laboratory, NASA Goddard Space Flight Center, Greenbelt, MD 20771, USA
\and
Department of Physics, University of Maryland, Baltimore County, 1000 Hilltop Circle, Baltimore, MD 21250, USA
\and
Columbia Astrophysics Lab 550 W120th St., 1027 Pupin Hall, MC 5247 Columbia University, New York, New York 10027, USA 
\and
Department of Physics and Astronomy, University of Pittsburgh, Pittsburgh, PA 15260
\and
University College London, Mullard Space Science Laboratory, Holmbury St. Mary, Dorking, RH5 6NT, U.K.
\and
INAF - Istituto di Astrofisica Spaziale e Fisica Cosmica, Via U. La Malfa 153, I-90146 Palermo, Italy
\and
Departamento de F\'isica y Astronom\'ia, Universidad de La Serena, Av. Cisternas 1200, La Serena, Chile.
\and
Instituto de Ciencias Astron\'omicas, de la Tierra y del Espacio (ICATE-CONICET), Av. Espa\~na Sur 1512, J5402DSP, San Juan, Argentina
 }

   \date{}

% \abstract{}{}{}{}{} 
% 5 {} token are mandatory
 
  \abstract{
 A sudden increase in the rate at which material reaches the most internal part of an accretion disk, i.e. the boundary layer, can change its structure dramatically. We have witnessed such change for the first time in the symbiotic recurrent nova T~CrB. Our analysis of \xmm, \swift\ Burst Alert Telescope (BAT)/ X-Ray Telescope (XRT) / UltraViolet Optical Telescope (UVOT) and American Association of Variable Stars Observers (AAVSO) V and B-band data indicates that during an optical brightening event that started in early 2014 ($\Delta$V$\approx$1.5):
% %   Munari mentions this also in page 7, where the comparison is with an spectrum taken in 2014-11
 ($i$) the hard X-ray emission as seen with BAT almost vanished; ($ii$) the XRT X-ray flux decreased significantly while the optical flux remained high;
%  reached a minimum after the optical peak; 
 ($iii$) the UV flux increased by at least a factor of 40 over the quiescent value; and ($iv$) the X-ray spectrum became much softer and a bright, new, blackbody-like component appeared. We suggest that the optical brightening event, which could be a similar event to that observed about 8 years before the most recent thermonuclear outburst in 1946, 
%  and was until now unexplained, 
 is due to a disk instability.
   }
\authorrunning{G. J. M. Luna et al.}
\titlerunning{Dramatic changes in T CrB. }
\maketitle
%-------------------------------------------------------------------

\section{Introduction} \label{sec:intro}

The interface between a Keplerian accretion disk and the accreting object is known as the boundary layer. This region, in a Keplerian disk, radiates approximately half of the available accretion luminosity, often in X-ray energies due to its high temperature. The accretion rate determines the optical depth of the boundary layer. 
A sudden change in the accretion rate can manisfest itself 
% is believed to be behind the dwarf novae outburst, which is inferred 
through a brightening in optical/UV and an X-ray fading.
Theory predicts a threshold above which the boundary layer will be optical thick to its own radiation and the observed spectrum will be blackbody-like, while below this threshold the spectrum will be that of an optically thin thermal plasma \citepads[e.g.,][]{1993Natur.362..820N,2014A&A...571A..55S}. 
% Observations however, point to different values for this threshold, as for example in the symbiotic star RT~Cru \citep{rtcru} or the ...,. \citep{}. 

T Coronae Borealis (T~CrB) is one of the four known recurrent novae where the companion is a red giant star, i.e. a symbiotic binary system. In these systems, strong eruptions are triggered by a thermonuclear runaway on the white dwarf (WD) surface after accretion of a critical amount of hydrogen-rich material from the companion. T~CrB had recorded nova-type outbursts in 1866 and 1946, when it reached magnitudes as bright as V=3, becoming a naked-eye object in the northern sky. It hosts a massive white dwarf, with M$_{WD}$=1.2-1.37 \ms\ \citepads{1998MNRAS.296...77B,2004A&A...415..609S}. Unlike in most symbiotics, in T~CrB the M4III \citepads{1999A&AS..137..473M} donor star fills its Roche-lobe \citepads{1998MNRAS.296...77B}, and accretion thus proceeds through the L1 point into an accretion disk before reaching the WD surface. The long orbital period of 227.5687$\pm$0.0099 days \citepads{2000AJ....119.1375F} implies an accretion disk that extends out to the circularisation radius R$_{circ}\approx$ 10$^{12}$ cm; i.e. a distance from the WD that has a Keplerian orbit with the same angular momentum that had the transfered material when passed through the Lagrangian internal point L1 \citepads[see eq. 5 in][]{2008ASPC..401...73W}. % \citepads[see eq. 5 in][]{2008ASPC..401...73W}.

In terms of its X-ray spectrum, T~CrB is one of five symbiotic stars with X-ray emission sometimes hard and bright enough to be detected with the \emph{Neil Gehrels Swift} Observatory Burst Alert Telescope \citepads[BAT;][]{2004ApJ...611.1005G,2009ApJ...701.1992K}. Observations of T~CrB with \rxte\ and \suzaku\ in 2006 and 2009 showed that the hard X-ray spectrum could be described by a hot, highly absorbed, optically thin thermal plasma from the accretion disk boundary layer \citepads{2008ASPC..401..342L,2016MNRAS.462.2695I}. \suzaku, \rxte\ and recent \nustar\ observations will be discussed in a forthcoming paper.
% The accretion rate derived from the spectral model was $\dot{M}\sim$4.2$\times$10$^{-9}$ \ms yr$^{-1}$ (d/1 kpc)$^{2}$.

The exhaustive historical optical light curve compiled by \citetads{2014AAS...22320901S} suggests 
% behavior that is unique among nova. 
that T~CrB brightened by about 1 mag in V several years ($\sim$ 8) before both recorded recurrent novae outburst. The origin and frequency of these brightening events is unknown, but they perhaps indicate a change in accretion flow onto the WD. Since the onset of the current optical brightening that started in early 2014, referred to as a "super-active" state, \citetads{2016NewA...47....7M} found that the luminosity of the ionizing source has increased, leading to a strengthening of high-ionization emission lines such as He~II $\lambda$4686\AA\ and the nebular radiation, which now overwhelms the red giant continuum.
% 1) the orbital  modulation in the B-band vanished; 2) high-ionization emission lines such as He~II $\lambda$4686\AA\ strengthened significantly, and the optical nebular radiation became so strong that it now overwhelms the red giant continuum; and 3) the luminosity of the ionizing source increased, leading to a larger ionization fraction of the red giant wind \citep{2016NewA...47....7M}. 
\citetads{2016NewA...47....7M}, however, only presented observations through December 2015. The light curve from the AAVSO indicates that the maximum brightness was reached around 4-5 April 2016 (see Fig. \ref{fig:aavso}). T~CrB was persistently detected with the \swift/BAT since \swift\ launch in late 2004, until late 2014 when 
% 
% the 14--50 keV flux started a 
% shallow 
% 
in the span of 4 100-day bins, the
BAT 14--50 keV flux declined from $\sim$4 mCrab to $\sim$2 mCrab, then exhibited a sudden drop to ~0 (within
1 sigma) in the following time bin.
% decline and between April, 2015, and April/June, 2016, the 14--50 keV \swift/BAT flux faded to an almost undetectable level.

% . Since April/June 2016, T~CrB has been barely detected with \swift/BAT.
 
In this article we study the super-active state, focusing on the \xmm\ observation taken about 300 days after the optical maximum.  Based on the behavior of the high-energy emission and the nature of the T~CrB system,
% " (where by 'nature of the T CrB system' we mean the fact that the RG fills its Roche lobe and the disk is so large that it is hard for it to be stable).
% of the high energy emission, 
we propose that the super-active state is due to a disk-instability. In Section \ref{sec:obs} we describe our dataset while in Section \ref{sec:result} we present the results from the spectral and timing analysis. Section \ref{sec:disc} presents our interpretation.

\section{Observations. \label{sec:obs}}

% {\bf what about here?}

\subsection{\swift}

 On January 18th 2017 we started a \swift~XRT+UVOT monitoring campaign, with observations first every week, then every two weeks, and during the last year, once every month. The \swift/XRT was operated in Photo Counting mode. We extracted source and background count rates from each observation
% spectra and light curves 
from a circular region with a radius of 20 pixels ($\approx$47$^{\prime\prime}$) 
% whose centroid we determined using the tool \texttt{xrtcentroid} and is xx$^{\prime\prime}$ away from 
centered on T~CrB SIMBAD coordinates ($\alpha$=15h 59m 30.16s; $\delta$=+25$^{\circ}$ 55$^{\prime}$ 12.6$^{\prime\prime}$). %15 59 30.1622 +25 55 12.613
We accounted for the presence of dead columns on the CCDs using the tool \texttt{xrtlccorr}. 
%  We extracted background source and light curves from an annular region with inner and outer radii of 25 and 40 pixels, respectively. We built the ancillary matrix (ARF) using the tool \texttt{xrtmkarf} and used the \texttt{swxpc0to12s0\_20010101v012.rmf} response matrix. 

\begin{figure*}[ht!]
 \begin{center}
\includegraphics[scale=0.8]{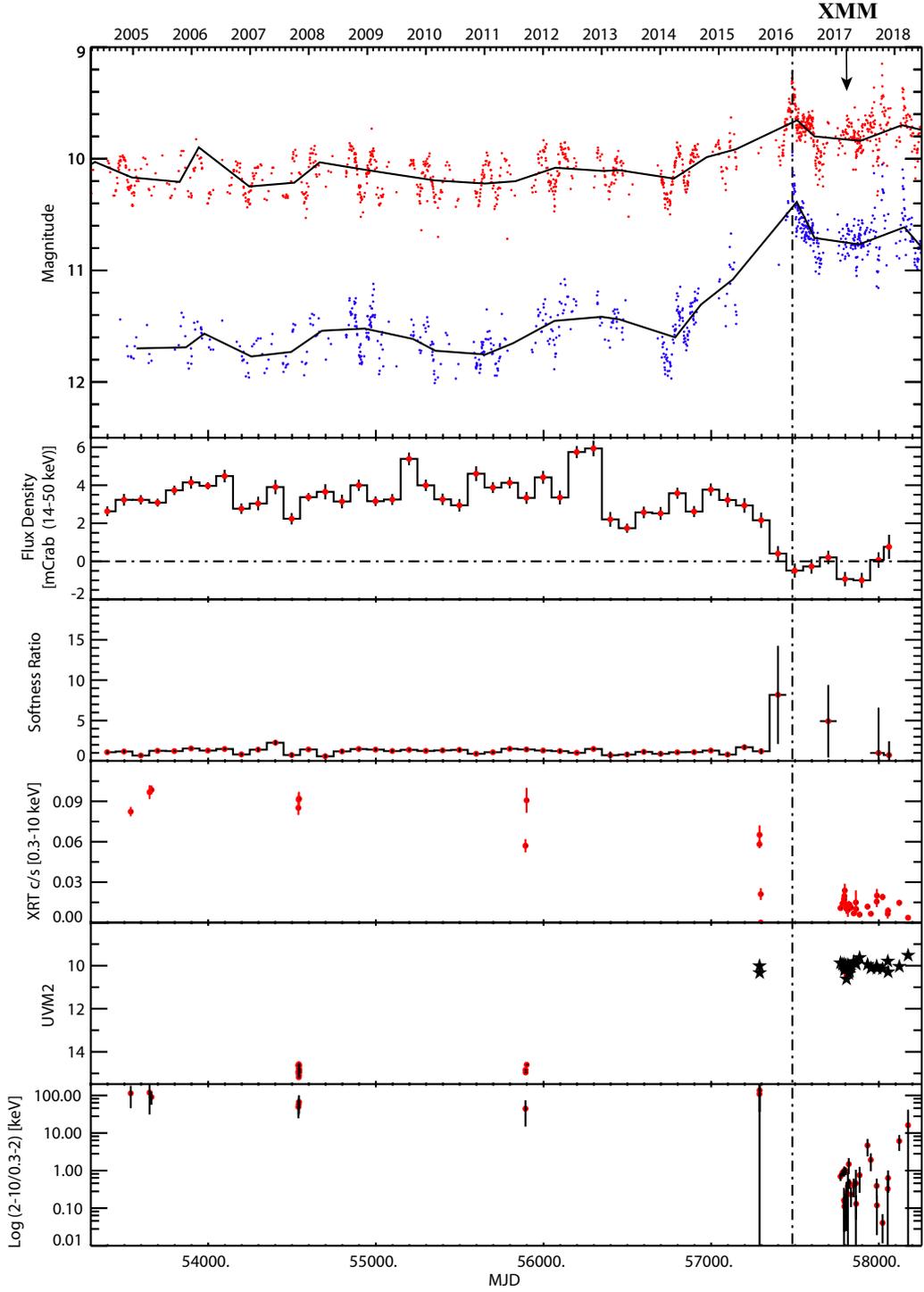}
\caption{{\em (a)}: AAVSO V- (red dots) and B-band (blue dots) light curves covering the period late 2004-early 2018. Vertical, dashed line shows the time of maximum optical brightness. 
% while the solid line marks the date of the \xmm\ observation and dashed red line shows the date of the \nustar\ observation reported in \citepads{luna18}. 
Black, solid line shows a curve from a moving average with 228 days (the orbital period) mean. {\em (b)} \swift\ BAT 14--50 keV light curve with 100 days bins. It is evident that the BAT flux started to decay quasi-simultaneously with the increase in the optical flux, which started around December 2014, it became too faint for its detection right around the optical maximum. {\em (c)} \swift\ BAT {\em softness} ratio (15--25/25--100 keV) with 100 days bins. This ratio steeply increased owing to the softening of the X-ray emission. {\em (d)} \swift\ XRT 0.3-10 keV count rate. 
% Since \swift\ started to observe T~CrB, there has been a clear decay in the X-ray flux, reaching its lowest level right after optical maximum. 
{\em (e)} \swift\ UVOT UVM2-magnitude light curve. Stars indicate magnitudes determined from the CCD readout streak. Even during the rise to optical maximum, the UV flux has already increase dramatically. {\em (f)} \swift\ XRT hardness ratio (2-10/0.3-2 keV). Since the beginning of the optical brightening, the X-ray emission has gotten about 100-times softer than before.  }
 \label{fig:aavso}
  \end{center}
 \end{figure*}

 \subsection{\xmm}

We observed T~CrB on 2017 February 23 (through a DDT time request) using the EPIC camera in Full Window mode, with the medium filter, for 53.8 ks and the OM camera in fast-mode. After removing intervals with high flaring background, the net exposure time reduced to 38.3 ks. Source, background spectra and light curves where extracted from circular regions of 32 and 42 arcsec radii, respectively, with the source region centered on T CrB SIMBAD coordinates and the background in a source-free region of the same CCD. We used the \texttt{RMFGEN} and \texttt{ARFGEN} to build the redistribution matrices and ancillary responses. The resulting X-ray spectra were grouped with a minimum of 25 counts per energy bin. For timing analysis, we converted photon arrival times to the Solar system barycenter using the SAS task \texttt{barycen}.
  
We emphasize that both grades distribution and offset maps indicate that the \swift\ and \xmm\ observations were not affected by optical loading on the X-ray detector, and the softest X-rays in the spectra are real. Very bright optical sources tend to create spurious X-rays photons, or change the grades and energies of X-ray photons in the case of moderately optically bright sources, and we have verified that this is not the case for T~CrB.

  \subsection{Optical/UV photometry.}

During each visit with \swift~ we also obtained UVOT exposures with the UVM2 ($\lambda$2246 \AA, FWHM=498 \AA) filter. The UVOT light curve was constructed using the \texttt{uvotproduct} tool. 
% Counts were converted to magnitudes and fluxes using the corresponding photometric zero points \footnote{\url{https://swift.gsfc.nasa.gov/analysis/uvot_digest/zeropts.html}}. 
% As can be seen in Figure \ref{fig:aavso}, 
Most UVOT observations taken during the current optical brightening are saturated with Vega-magnitudes brighter than about 10. However, we were able to measure source magnitudes using the readout streak, as detailed by \citetads{2013MNRAS.436.1684P}.
Observations with the Optical Monitor (OM) onboard \xmm\, with the V, U, B, W1 filters were also saturated.

We also collected multi-epoch photometric observations in the V and B bands from the American Association of Variable Star Observers (AAVSO) 
% and the "All Sky Automated Survey" \citep[ASAS]{2002AcA....52..397P} 
to study the X-ray data in the context of the optical state.
%  and the American Association of Variable Star Observers (AAVSO). 
The observations covered a period of $\sim$4200 days.
% , since T$_{0}$= JD2400000.5. 
% For the ASAS data, we only considered those with quality flag "A" in the GRADE column and aperture to 30 arcsec. In the case of AAVSO we ignored those data without measurement error bar. 

% % % % % % % % % % 
% %  simultaneouly Swift and Nustar observations, 2015-09-23 04:46:08, obsid 00081659001
    
\section{Analysis and results. \label{sec:result}}

% Between late 2014, 
%  fade until being almost undetectable since April/June 2016;
The XRT flux reached its lowest value since the launch of \swift\ during our observing campaign that started in 2017 January.
%  the 0.3-10 keV \swift/XRT flux has been slowly decaying since late 2004, reaching the actual low flux value during our campaign that started in January 2017. 
Along with the X-ray fading, the XRT spectra have clearly softened, with a soft component dominating the spectra at energies less than about 1 keV. T~CrB increased its brightness in UV significantly, from about 15 to brighter than approximately 9.8 UVM2 mag. The upper panel in Figure \ref{fig:aavso} shows the resulting optical, X-ray and UV light curves.

\begin{figure*}[ht!]
 \begin{center}
\includegraphics[scale=0.9]{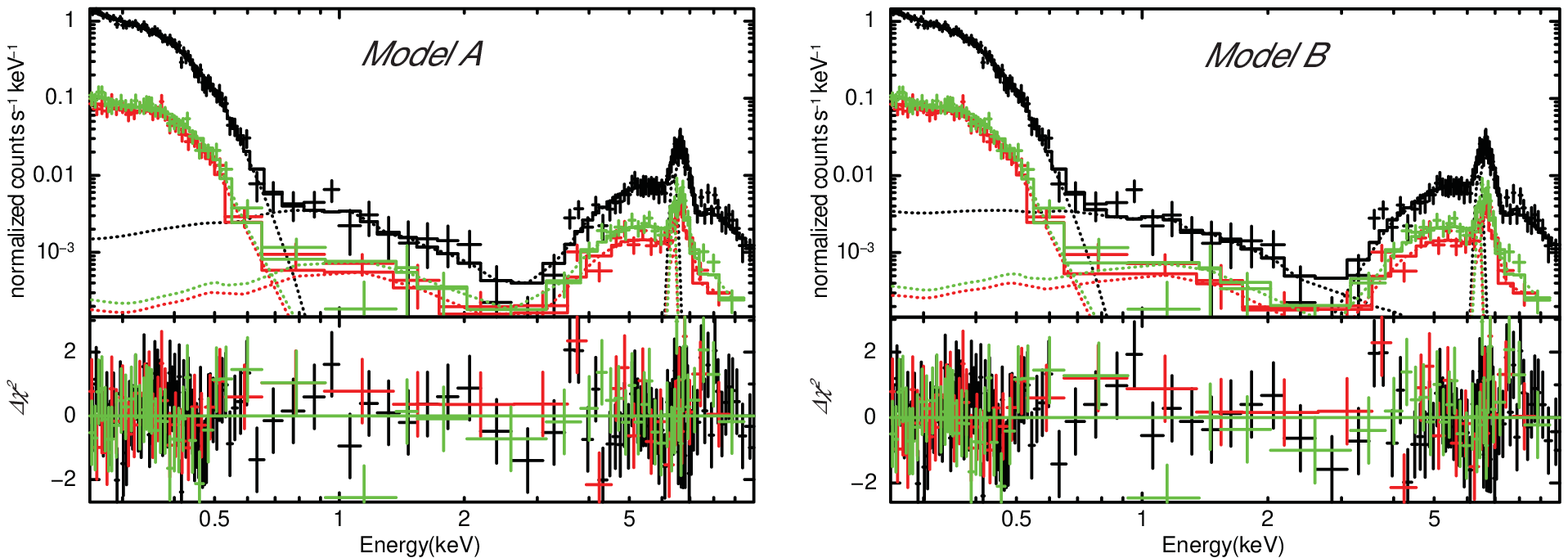}
\caption{{\em Left:} \xmm~ $pn$ (black), MOS1 (red) and MOS2 (green) X-ray spectra of T~CrB  data with the best spectral model (solid line) and the contribution of each model' component (dotted lines). The model consists of a blackbody plus an optically thin cooling-flow and a Gaussian profile centered at the Fe K$\alpha$ energy of 6.4 keV. All components are absorbed by interstellar and local, partial covering absorption (see Table \ref{tab:fits} for resulting parameters). 
{\em Right:} An alternative model, with a soft blackbody, a hard thermal plasma component, and a medium energy optically thin thermal component to take into account photons in the $\sim$0.6--3 keV region. The soft optically thick and medium energy, optically thin components are modified by mostly interstellar absorption, while the hard, cooling flow is also affected by a partial covering absorber. Both bottom panels show the fit residuals in units of \chisq. }
 \label{fig:xmm}
 \end{center}
 \end{figure*}

On 2017 February 23, well into the super-active state, our deep observation with \xmm\ showed that the EPIC spectra (Fig. \ref{fig:xmm}) are obviously complex, and can be divided into three energy ranges. Above 3 keV, there is a highly absorbed component with prominent emission line complex in the 6--7 keV range, indicative of optically thin, thermal origin. This is likely the same $\delta$-type component seen in T~CrB in its normal state \citepads{2009ApJ...701.1992K,2013A&A...559A...6L}.
% , so we model this with a cooling flow model with an additional Guassian line at 6.4 keV. 
Below $\sim$0.7 keV, the spectra are dominated by a soft, unabsorbed component. A blackbody provide an good description of this region. Photons are also detected in the intermediate energy range (0.7--3 keV). We fit the spectra in two different ways. 

% In one, the entire 0.6-10 keV spectra are assumed to be the $\delta$-type component seen through a partial covering absorber 

In one, all components are absorbed by interstellar absorption and local, partial covering absorption that blocks 99.7\% of the emission (model A in Table \ref{tab:fits}, see also Fig. \ref{fig:xmm}). The soft X-ray spectra consist of blackbody-like emission (with T$_{bb}$=4$\times$10$^{5}$ K) from a region smaller than the surface of the WD, with a spherical surface area of 4.2$\times$10$^{7}$ km$^{2}$ (in the case of a WD with 1.2 \ms\ and R$_{WD}$=5$\times$10$^{8}$ cm, this represents approximately 13\% of the surface of the WD). The hard (0.6$\lesssim$ E $\lesssim$ 10 keV) spectra were consistent with a multi-temperature, cooling flow, with maximum temperature kT$_{max}$=12.9 keV. An alternative to this two-components model, where only the $\delta$-component is partially-covered by the absorber yielded a blackbody-component spherical surface area of 1.5$\times$10$^{5}$ km$^{2}$ (model A').

% bbodyrad_mkcflow_apec.xcm
We also studied an alternative model, with three separate thermal components, a soft blackbody-like component, a hard optically thin thermal plasma component and a medium energy optically thin thermal plasma component to take into account photons in the $\sim$0.7--3 keV region. 
% The addition of this third component is motivated phenomenologically rather than physically. 
The soft, optically thick and medium-energy, optically thin components are modified by mostly interstellar absorption while the hard, cooling flow is affected by full and partial covering absorbers. Although in terms of \chisq (0.95/222 d.o.f) this model is also acceptable, parameters such as the absorbing column densities or the covering fraction, are unconstrained. The black-body emitting area in this case was smaller than in model A, with a value of 1.3$\times$10$^{5}$ km$^{2}$ (covering much less than 0.1\% of the WD surface; model B in Table \ref{tab:fits}).  

\begin{figure}[ht!]
%  \begin{center}
\includegraphics[scale=0.5]{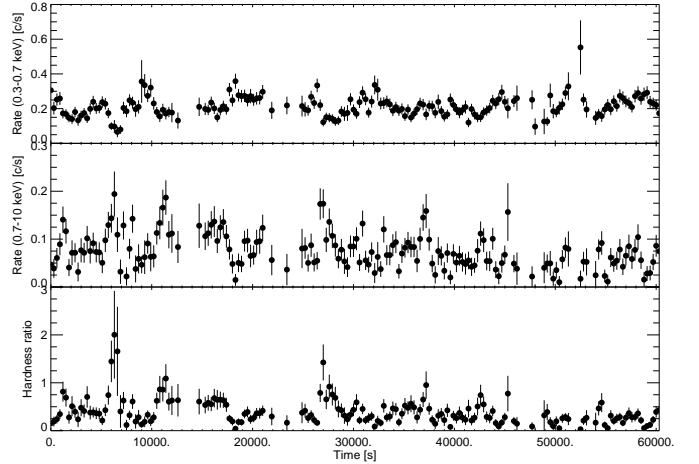}
\caption{Background-subtracted \xmm\ EPIC light curves with bins of 300 s. {\em Top}: 0.3--0.7 keV. {\em Middle}: 0.7--10 keV. {\em Bottom}: Hardness ratio (0.7--10 keV/0.3--0.7 keV)}
 \label{fig:hr}
%   \end{center}
 \end{figure}

The light curves from the \xmm\ observation, binned at 300 s, show strong variability with fractional amplitudes of 0.32 in the soft (0.3--0.7 keV) X-ray band and 0.47 in the hard (0.7--10 keV) X-ray band, respectively (Figure \ref{fig:hr}).  The \xmm\ X-ray light curves 
% revealed variability on a time scale of 300 s, it 
did not contain statistically significant periodic modulations. We searched for periodicities in the X-ray light curves (with bins of 300 s) in the 0.3--0.7 and 0.7--10 keV energy ranges by computing the Fourier power spectrum. 
% and fit a simple power law to the region dominated by red noise, 
The log-log power spectrum in the frequency region $f~\lesssim$ 0.0033 Hz is dominated by red noise.
% whereas for $f~\gtrsim$ 0.001 Hz, the light curve was dominated by Poisson counting statistic. 
To search for periods with amplitudes in excess of the red-noise,
% dominated frequencies, 
we 
% thus extracted new light curves with bins of 1000 s and 
modeled the log-log power spectrum with a simple power-law using a Least-squared fit \citepads{2005A&A...431..391V}. The detection threshold was determined by simply assuming that the model is a good description of the underlying power spectrum and thus their ratio will be distributed like $\chi^{2}$. We estimated the probability (at a 95 and 99.74\% confidence levels) that at a given frequency, a large peak would be present in the periodogram by comparing the ratio of modeled over observed power spectrum to the $\chi^{2}$ probability distribution.
No peaks in the power spectrum excedeed the detection threshold. 
% in the Poisson and red-noise regions 
None of the power spectra, in any of the selected energy ranges, show periods with a significance greater than 3$\sigma$. Our observations were sensitive to pulsed fraction \citepads[following][]{1996ApJ...468..369I} greater than 15\% for $f~\lesssim$ 0.0033 Hz in the 0.3-0.7 keV range and greater than 22\% for $f~\lesssim$ 0.0033 Hz in the 0.7-10 keV range.

% % %  OM brightness limit
% %  http://www.mssl.ucl.ac.uk/www_xmm/ukos/onlines/uhb/XMM_UHB/node70.html

% \begin{figure}%f1
%  \includegraphics[scale=0.5]{../../XMM/dataprep/lc_b1000_sss_250_700.ps}
%  \caption{}
%  \label{fig:psd}
%  \end{figure}

% % % % % % % % % % % %  Portrait table
\renewcommand{\arraystretch}{1.3}
\begin{table}
\caption{X-ray spectral fitting results. Unabsorbed X-ray flux and luminosity, in units of 10$^{-13}$ ergs s$^{-1}$ cm$^{-2}$ and 10$^{31}$ ergs s$^{-1}$, respectively, are calculated in the 0.3-50 keV energy band for the optically thin thermal components and in the 1 eV to 10 keV for the optically thick thermal component. Elemental abundances are quoted in units of abundances from \citet{2000ApJ...542..914W}. Luminosity, size of the black-body emitting region, and $\dot{M}$ are determined assuming a distance of 806 pc. \label{tab:fits} } 
\centering
\begin{tabular}{lccccc}
\hline\hline
Parameter & Model A$^{1}$ & Model B$^{2}$ \\
\hline
$N_{H}^{a}$ [10$^{22}$ cm$^{-2}$] & 0.049$\pm$0.002 & 63$_{-34}^{+12}$	\\
$N_{H}^{b}$ [10$^{22}$ cm$^{-2}$]& 68$\pm$2 & $\gtrsim$39\\
CF$^{b}$ & 0.997$\pm$0.001 & $\gtrsim$0.6 \\
kT$_{max}$ [keV] & 12.9$\pm$0.5 & 13$\pm$5	\\
Z/Z$_{\odot}$ & 1.5$\pm$0.3  & 1.5$_{-0.6}^{+0.8}$	\\
$\dot{M}$ [10$^{-9}$ M$_{\odot}$ yr$^{-1}$] & 0.16$\pm$0.01 & 0.17$\pm$0.1	\\
$N_{H}^{c}$ [10$^{22}$ cm$^{-2}$]& \nodata & $\lesssim$0.09\\
kT$_{mid}$ [keV] & \nodata & 4.5$_{-2}^{+25}$	\\
$N_{H}^{d}$ [10$^{22}$ cm$^{-2}$]& \nodata & 0.05$\pm$0.01	\\
kT$_{bb}$ [keV] & 0.035$\pm$0.001 &0.035$\pm$0.002\\
R$_{bb}$  [km] & 1830$\pm$30& 103$\pm$10\\
F$_{cf}$ & 61.0$\pm$0.2  &  65$\pm$2	\\
F$_{mid}$ & \nodata  & 0.15$\pm$0.3	\\
F$_{bb}$ & 86000$\pm$1000  & 285$\pm$5	\\
L$_{cf}$ & 47 &	50.5\\
L$_{mid}$ & \nodata & 0.12\\
L$_{bb}$ & 66800 & 221\\

$\chi^{2}_{\nu}$/dof & 0.92/226& 0.95/222\\
% & 10$^{22}$ cm$^{-2}$ & & keV & [km] & keV  &  & 10$^{-9}$ M$_{\odot}$ yr$^{-1}$ & & &\\
% HJD & $E$ & Method\#2 & Method\#3 \\    % table heading 
\hline
\end{tabular}
\tablefoottext{1}{\texttt{TBabs$^{a}\times$(partcov$\times$TBabs)$^{b}\times$(bbodyrad + mkcflow + gaussian)}} \\
\tablefoottext{2}{\texttt{(TBabs$^{a}\times$(partcov$\times$TBabs)$^{b}\times$(mkcflow + gaussian)+TBabs$^{c}\times$apec+TBabs$^{d}\times$bbodyrad)}}
\end{table}

% % % %  Fe Ka line EW 190 eV

\section{Discussion. \label{sec:disc}}

% % %  NO OPTICAL LOADING; see pnspec_gr0.fits and pnspec_gr1.fits

% However, the resulting parameters are difficult to interpret physically. For example, an extra optically thin thermal component, with a temperature of $\sim$5 keV, is necessary to account for the photons in the $\sim$1 to $\sim$ 3 keV energy range but this temperature is too high to attribute its origin to for example, colliding winds, which would imply shock at speeds of xxxx km s$^{-1}$. Such speeds ae not observed in symbiotics or recurrent novae. Additionally, the size of the blackbody-type emitting region seems too small, about 120 km. The lack of pulsations in the X-ray light curves makes it difficult to interpret this as emission from the polar caps of a magnetic accretion column.
\subsection{The boundary layer and unstable disk}

%1%
From the spectral models fit to the \xmm\ EPIC spectra, we consider model A to be the most compelling. While the size of the blackbody-component in models A' and B are unlikely small to arise from the WD surface or the accretion disk boundary layer, in model A it might well corresponds to the size of the boundary layer.
Changes in strength and character of UV and X-rays, along with optical changes reported by \citetads{2016NewA...47....7M,2016ATel.8675....1Z,2016MNRAS.462.2695I}, demonstrate that after early 2014, the rate of mass transfer onto the WD, $\dot{M}$ increased and the boundary layer became predominantly optically thick. The non-detection of periodicities in the light curves of both the hard and soft X-rays suggests that the emission is not powered by magnetically-channeled accretion. The stochastic variability in the hard band is, however, expected if the emission arise in the accretion disk boundary layer. Furthermore, the strong variability observed in the soft band and the small area of the blackbody emitting region derived from the spectral model A
% supports origin of the X-ray emission in the accretion disk boundary layer instead of 
% Whereas the 
% , the presence of flickering in this same energy range and 
% the small area of the emitting region derived from the spectral model
indicates that this emission is not powered by thermonuclear burning on the WD surface.
% or the base of a magnetic accretion column. 
% The normalization from the spectral model of the super-soft component provides the area of the emitting region. If the region is indeed the boundary layer and it has a ring shape, 

At a distance of 806$^{+33}_{-30}$ pc \citepads[as recently determined with $GAIA$][]{2018arXiv180410121B},  
% a ring-shape boundary layer, the viewed area ($A$) varies with the $\cos{i}$, where $i$=67 deg \citep{2004A&A...415..609S}, 
% thus its 
the unabsorbed luminosity of the blackbody-emitting region
% of the super-soft component 
is L$_{bb}$=$A\sigma$T$^{4}$=6.68$\times$10$^{35}$ (d/806 pc)$^{2}$ \lumcgs, with A being the surface area of a sphere with radius R$_{bb}$=1.8$\times$10$^{8}$ cm. The UVM2 fluxes indicate that 
% the luminosity 
L$_{UV} > $ 2$\times$10$^{34}$ (d/806 pc)$^{2}$ \lumcgs. Assuming that half of the accretion luminosity is radiated in the boundary layer, and that M$_{WD}$=1.2 \ms and R$_{WD}$=5$\times$10$^{8}$ cm, 
% and that this luminosity is the sum of the luminosities from the optically thin and thick portions, 
we found that the accretion rate feeding the optically thick portion of the boundary layer was $\dot{M}_{bb}\approx$6.6$\times$10$^{-8}$ (d/806 pc)$^{2}$ \ms\ yr$^{-1}$.
% then we estimate the outer radius of the boundary layer to be R$_{L} \sim$0.11 R$_{WD}$ in the case of model A and khskjchns for model B. In each case, the corresponding luminosity is 3$\times$10$^{34}$ erg s$^{-1}$ or xcscs$\times$10$^{34}$ erg s$^{-1}$, respectively. 
The luminosity of the boundary layer is the sum of the luminosity of the optically thick and optically thin cooling flow components, L$_{BL}$=L$_{bb}$+L$_{cf}$. The luminosity of the optically thin portion of the boundary layer, from the cooling flow spectral model in the 0.3-50 keV, was L$_{cf}\approx$4.7$\times$10$^{32}$ (d/806 pc)$^{2}$ \lumcgs. The accretion rate feeding the optically thin portion of the boundary layer was thus $\dot{M}_{cf}$=1.6$\times$10$^{-10}$ (d/806 pc)$^{2}$ \ms~ yr$^{-1}$.
% feeding the optically thick portion is $\dot{M}_{bb}$=3.2$\times$10$^{-8}$ \ms\ yr$^{-1}$.
% , with 3.9$\times$10$^{-9}$ \ms yr$^{-1}$ corresponding to the optically thick portion. 
These accretion rates are consistent with the expected theoretical values where the boundary layer is optically thick/thin to its own radiation \citepads{1985ApJ...292..535P,1993Natur.362..820N}.

% For the first time, there is direct evidence of an outburst driven by an increase on $\dot{M}$ that changed the optical depth of the boundary layer in a symbiotic binary. For the first time also, we have unambiguously identified such disk-instability in a recurrent nova.  
% (right? for all the other RNe) 

The current high optical brightness state appears to be associated with the highest accretion rate on record.
The long term optical light curve of T~CrB shows a history of different levels of activity \citepads[see][]{2004A&A...415..609S,2016NewA...47....7M,2016MNRAS.462.2695I} and with the exception of the nova eruptions, all of them had lower intensity than the current level. We can rescale, at $d$=806 pc, the reported accretion rates and compare then with the current level. The U-band light curve 
% In their analysis of optical and IUE data from May 1979 until March 2001,  found that 
from May 1979 until August 2002 presented by \citetads{2004A&A...415..609S} shows that T~CrB was in a low brightness state from JD 2447300 -- JD 2450000 (1987 August through 1995 September), while from JD2444300 to JD2447300 (1980 March through May 1988) was in an optical high state. By modeling the IUE spectra during the low and high states, \citetads{2004A&A...415..609S} found $\dot{M}_{low}$=1.53$\times$10$^{-9}$ (d/806 pc)$^{2}$ \ms~yr$^{-1}$ and $\dot{M}_{high}$=1.1$\times$10$^{-8}$ (d/806 pc)$^{2}$ \ms~yr$^{-1}$. 
\citetads{1992ApJ...393..289S} analyzed IUE data and suggests that during the decade of 1980, the accretion rate was on average $\dot{M}_{low}$=9.6$\times$10$^{-9}$ (d/806 pc)$^{2}$ \ms~yr$^{-1}$. 
% Thus we conclude that a relatively small increase in $\dot{M}$ has caused 
The optical depth of the boundary layer 
% during these past periods, however, is unknown. None of the 
previous to \swift\ launch is unknown. 
% X-ray sky survey
% detected T~CrB. 
Only short pointed X-ray observation of T~CrB ($\approx$2 ks) with {\em Einstein} obtained on 1979-02-26 were reported by \citetads{1981ApJ...245..609C}, with a 
% As for the EInstein observation: The observation was obtained on 1979-02-26 and Cordova et al.
% 1981 (ApJ 245, 609) reported a count rate of 0.0083+/-0.003 c/s with IPC, and a 
0.1--4.5 keV luminosity of 7.2$\times$10$^{30}$ (d/806 pc)$^{2}$ erg/s. In this energy range, T~CrB was even fainter than the current X-ray state. It is unknown, however, if a hard X-ray component existed in the past. 
% Although the current optically active state is unprecedented in its strength, T CrB's low X-ray flux during the nineteen eighties suggests that the boundary layer might also have been optically thick then

% The same data are listed as 4.4e31 erg/s in the Cordova & Mason 1984
% (MNRAS 206, 879) paper. (Hmm, needs to check the assumed distance for both papers.) 
% Either way, all we can say is that it wasn’t  obviously luminous, but if T CrB was strongly absorbed
% in 1979 February, there could have been a strong hard component.

The current brightness state, with $\dot{M} \gtrsim$6.6$\times$10$^{-8}$ (d/806 pc)$^{2}$ \ms yr$^{-1}$ being the highest on record, dramatically changed the structure of the boundary layer, making it mostly thick to its own radiation.
% , was thus caused by a unnoted increase in $\dot{M}$. 
\nustar\ observations in Luna et al. (in prep.) support this interpretation.
% \nustar\ data analyzed in a forthcoming article are well described by an absorbed, multi-temperature plasma, with $\dot{M}$=2$\times$10$^{-9}$ \ms yr$^{-1}$. This observation was taken well on the rise of the optical brightness (see Fig. \ref{fig:aavso}). A simultaneous \swift/UVOT/XRT observation allows us to measure the ratio of UV to X-ray fluxes, which can be used as proxy for the optical depth of the boundary layer \citepads{2018arXiv180102492L}. The ratio F$_{UV}$/F$_{X}$ was greater than 3 (even withouth accounting for the unknown reddening). This ratio is clearly on the regime of a mostly optically thick boundary layer.  
% In T~CrB, the transition from mostly optically thin to mostly optically thick must occur when $\dot{M}$ is on the narrow range of 1--3$\times$10$^{-8}$ \ms yr$^{-1}$. 
Theory predicts that the transition from mostly optically thin to mostly optically thick should occur at accretion rates of about $10^{-9}$ \ms yr$^{-1}$ for a 1 M$_{\odot}$ WD \citepads{1995ApJ...442..337P} or even lower values as proposed by \citetads{2014A&A...571A..55S}. The accretion rate before the optical brightening was therefore likely smaller than a few $10^{-9}$ \ms yr$^{-1}$.

% Recently we have found that this predicted threshold does not apply to the massive WD accreting in the symbiotic system RT~Cru \citep{rtcru}. If the same applies to the boundary layer around the massive WD in T~CrB, it thus seems that a new theoretical determination of the accretion rate where the transition from optically thin to thick occurs is necessary.
% in the boundary layer around a massive WD ($M_{WD} \gtrsim$ 1 \ms).

Phenomenologically, the simultaneous quenching of the hard X-rays, with the remarkable increase in UV flux is exactly what is observed in well-known dwarf novae in outburst. Also, as it is observed in some dwarf novae \citepads{2003MNRAS.345...49W}, in T~CrB the maximum temperature of the cooling flow dropped from kT$_{max}$=57$\pm$10 keV as observed with \suzaku\ in September 2006 \citepads{2008ASPC..401..342L} to kT$_{max}$=12.9$\pm$0.5 keV during the \xmm\ observation reported here. The origin of the residual hard X-ray emission during outburst and the decrease of kT$_{max}$, are still a matter of debate and could be related to an accretion disk coronae or the presence of an additional cooling mechanism \citepads[see][for a detailed discussion]{2017PASP..129f2001M}.

% The enhancement of the accretion rate in the accretion disk must be due to an increased mass transfer rate from the donor, a disk instability, or both. Although illumination of the red giant could increase the rate of mass transfer into the accretion disk and illumination effects have been inferred from the IR light curves in quiescence \citepads{2016NewA...47....7M}, the hard X-ray emission that could reach the red giant photosphere and enhance the red giant mass loss rate during this high-state is too weak, i.e. a few orders of magnitude fainter than the red giant luminosity, making almost impossible for hard X-ray photons to penetrate the red giant' photosphere.

The enhancement of accretion rate in the accretion disk must be due to an increased mass transfer rate from the donor, a disk instability, or both. Although irradiation is cited as a potential cause of enhanced mass transfer in some interacting binaries, it is unlikely to explain the super high state of T~CrB. First, the irradiating flux is low:  \citepads{2016NewA...47....7M} used the infrared light curves to infer a temperature increase of the heated side of 80 K, or about 2\% increase compared to the unirradiated side, implying an irradiating flux of $\sim$8\% of the intrinsic stellar flux. Second, \citetads{2004A&A...423..281B} concluded that irradiation of a giant donor cannot result to mass transfer cycles. Finally, the time scale on which the envelope of the giant responds to changing irradiation is many thousands of years according to the same authors.

% \citepads[see,][and references therein]{2015AcPPP...2..246M}
% , illumination of the red giant could increase the rate of mass transfer into the accretion disk affecting the progression and/or duration of the active state. However, the illumination of the the red giant' facing hemisphere by the white dwarf/accretion disk cannot trigger a sudden increase in the accretion rate. 

In a steady-state disk, the effective temperature as a function of disk radius is given by \( T(R)=(3G/8\sigma\pi)^{1/4}\dot{M}^{1/4} M_{WD}^{1/4} R^{-3/4} \) \citepads[eq. 5.43 in][ with R $>>$ R$_{WD}$]{2002apa..book.....F} and has to be greater than 10$^{4}$ K, enough to keep H ionized. In T~CrB, with a WD mass of 1.2 \ms~ and a mass accretion rate $\dot{M}\sim$6.6$\times$10$^{-8}$ \ms~yr$^{-1}$ (determined from the \xmm\ observation), the steady-state disk would extend out to about R$\approx$0.75 R$_{\odot}$. Given the size of circularisation radius in T~CrB, it is highly unlikely that the entire disk can remain in a high state (with T$\gtrsim$10$^{4}$ K). If the region in the vecinity of the circularisation radius is cold, and unstable with mostly neutral H and low viscosity, then the accreted matter would accumulate near that region until a disk instability develops, allowing the material to flow inwards. 
% While a similar consideration can apply to RS~Oph and other systems, if they are wind accretors, the details would be different. 
A similar inside-stable, outside-unstable, hybrid accretion disk has been discussed for the recurrent novae RS~Oph by \citetads{2008ASPC..401...73W} in an attempt to explain how mass is transferred during quiescence periods, given the low accretion rate determined from X-ray observations in quiescence \citepads{2011ApJ...737....7N}. In this scenario, disk instability outbursts in the outer parts of the disk are infrequent, with recurrence time of the order of hundreds of years. Recently, \citetads{2018arXiv180407916B} studied a similar scenario, where the amount of material necessary to trigger a thermonuclear outburst in \object{RS~Oph} every $\sim$20 years is delivered to the WD in a series of disk-instability outbursts, with optical brightening amplitudes of about 1 magnitude. A similar scenario could be at work in \object{T~CrB}.

\subsection{X-ray absorption} 

%2 
% With an orbital period of 227.55 days \citep{2000AJ....119.1375F} and a giant that fills its Roche-lobe, the accretion disk that surrounds the WD, is thermally unstable (check) \citep{2011MNRAS.418.2576A}
X-ray observations from the past decade indicate that the absorbing column was very high during that time 
% that the absorption column toward the X-ray emitting material is normally very high 
\citepads{2008ASPC..401..342L,2009ApJ...701.1992K,2016MNRAS.462.2695I}, completely covering the source (Luna et al., in prep.) and absorbing all X-rays with energies lower than $\sim$2-3 keV. This might lead us to suspect that the blackbody-like spectral component could have been permanently emitting at the current level but was only detected now that there was a favorable sight. However, optical spectra during the past decade \citepads[see][]{2016NewA...47....7M} indicate that, with the exception of the current super-active state, the intensity of highly-ionized emission lines, such as HeII$\lambda$4686 has been low and the optical continuum weak, in contrast to what is observed now in the super-active state, simultaneous with the appearance of the blackbody-like component. If blackbody-like X-ray emission had been present but absorbed, we would have expected very strong optical line and continuum emission from the absorbing material.

Partial-covering absorber is sometimes inferred for $\delta$-type symbiotics. 
% (that includes the uzaku observation of T CrB, by the way, at least according to my analysis). 
However, a covering fraction of 99.7\% is extreme. While we cannot exclude a geometrical explanation for this
covering fraction, we have also explored a possible alternative. In high accretion rate, non-magnetic CVs, accretion disk wind features are routinely observed in the UV spectra, and perhaps also in the optical spectra \citepads{2015MNRAS.450.3331M}. It therefore makes sense to ask if the accretion disk wind could be responsible for the partial covering absorption we observe in T~CrB in high-$\dot{M}$ state in particular, and in $\delta$-type symbiotic stars in general. Matthews and others consider a wind mass loss rate of 10$^{-9}$ M$_\odot$ yr$^{-1}$ to be a reasonable estimate for an accretion rate of 10$^{-8}$ M$_\odot$ yr$^{-1}$; the former is a non-linear function of the latter, such that the fraction of mass lost in the wind is lower for lower accretion rate. Their model (see a schematic in Figure 3 of Matthews et al.) has a biconical geometry with an X-shaped cross section, with wind being launched from 4--12 R$_{wd}$. The launch radii may be particularly uncertain as wind is launched from the innermost regions of the disk in other simulations \citepads[e.g.,][]{2018MNRAS.475.3786D}.
% We have performed the following order-of-magnitude estimate of the resulting absorbing column. 
We express the mass loss rate in the units of 10$^{-9}$ M$_\odot$\,yr$^{-1}$ as $\dot M_{w,-9}$; for $\dot M_{w,-9}$=1.0, the mass loss rate is $2.0 \times 10^{24}$ g\,yr$^{-1}$=$6.3 \times 10^{16}$ g\,s$^{-1}$. We further express the characteristic radius of the wind where the line-of-sight to the X-ray emitting region crosses it in the units of 10$^9$ cm as $r_{w,9}$, and the wind velocity at the same point in the units of 1000 km\,s$^{-1}$ as $v_{w,8}$.  Dividing mass loss rate by 4$\pi$r$v$ , for $r_{w,9}$=1.0 and
$v_{w,8}$=1.0, we obtain a mass column density expected from a wind of $5.0 \times 10^{-2}$ g\,cm$^{-2}$, or $N_H^{wind} \sim 5 \times 10^{22}$ cm$^{-2}$.

Given our estimate for the accretion rate in T~CrB in the high state of $6.6 \times 10^{-8}$ M$_\odot$\,yr$^{-1}$, $\dot M_{w,-9}$=3.0 would be a plausible assumption. According to \citetads{2015MNRAS.450.3331M}, it takes the wind of order 100 R$_{wd}$ to obtain its final velocity, and it has poloidal velocities of less than 100 km\,s$^{-1}$ within a few R$_{wd}$, so we can try $v_{w,8}$=0.1.
% (the wind does retain high rotational velocity of the Keplerian accretion disk, but that does not affect our calculation here.) 
Then $r_{w,9}$=2.2 would result in the observed $N_H$ value of $6.8 \times 10^{23}$ cm$^{-2}$. This admittedly crude estimate suggests that the accretion disk wind is a plausible origin of the local absorber in T CrB in high state. 

However, our Model A suggests a partial covering absorber with a covering fraction of 99.7\%, and the remaining $\sim$0.3\% experiences hardly any absorption at all ($4.5\times 10^{20}$ cm$^{-2}$, or less than a thousandth of the column of the partial-covering absorber, and this may well be dominated by the ISM). While the 3D simulation of \citetads{2018MNRAS.475.3786D} does show a clumpy wind, the density contrast between the clumps and inter-clump regions is only of order 10. In general (i.e., regardless of whether the absorber is the accretion disk wind or something else), we do have not identified a natural explanation for a near 100\% covering fraction and a factor of 1000 density contrast.

The scattering model for CH Cyg of \citetads{2006MNRAS.372.1602W} may provide a possible alternative to our best-fit spectral model with its extraordinarily high absorption covering fraction. This model requires a low density matter that gets highly ionized, and a clear line of sight to that region. Such a region could exist in the polar cavity of the wind. For this region to remain highly ionized, it cannot be optically thick to X-rays from the central source, which also limits its scattering efficiency. Combined with geometrical factors, it appears plausible for this region to scatter $\sim$0.3\% of the total luminosity into our line of sight.

In a normal state there are problems for the accretion disk wind as the origin of X-ray absorbers in all $\delta$-type symbiotic stars. One is that we would expect a strong inclination angle dependence, with little absorption when one is seen pole-on. The other is the accretion rate dependence of the wind mass loss rate. The numbers appear to work for T CrB in the high state. In a normal state, the accretion rate may only be a few times 10$^{-9}$ M$_\odot$\,yr$^{-1}$, hence the wind mass loss rate is expected to be of order 10$^{-10}$ M$_\odot$\,yr$^{-1}$ at most. Yet $N_H$ in excess of $10^{23}$ cm$^{-2}$ is seen in T CrB in a normal state. If all X-ray absorbers in $\delta$-type symbiotics are due to accretion disk wind, we would expect a stronger dependence on the accretion rate. Similarly, this model may have difficulty explaining the high $N_H$ values seen in lower X-ray luminosity $\delta$-type systems.

% Although both spectral models are statistically acceptable, we prefer model A as the best fit model given that model B give parameters that seems implausible. The blackbody emitting area implied by model B is too small, on the order or 1.7$\times$10$^{4}$ Km$^{2}$. And given the lack of periodicities in the X-ray light curves, it is difficult to picture a scenario where an small portion of the boundary layer is hot enough to emit soft X-rays 

At the time of submission, the current optically bright state continues. Our finding that the accretion-disk boundary layer around the $>$ 1.2 \ms WD in T~CrB transitioned from primarily optically thin to primarily optically thick when the accretion rate rose from $\sim$10$^{-9}$ \ms yr$^{-1}$ to $\sim$10$^{-8}$ \ms yr$^{-1}$ (assuming d=806 pc) challenges theoretical boundary-layer models, which must also explain BL transitions at much lower accretion rates for some DNe in outburst.

\begin{acknowledgements}
Based on observations obtained with \xmm\, an ESA science mission with instruments and contributions directly funded by ESA Member States and NASA.
We thanks the entire \swift\ team for accepting and planning our multiple Target-of-Opportunity requests. We thanks Norbert Schartel for approving the \xmm\ DDT request. We acknowledge with thanks the variable star observations from the AAVSO International Database contributed by observers worldwide and used in this research. We thanks Ulisse Munari for the helpful discussions about the current super-active state. GJML and NEN are members of the CIC-CONICET (Argentina) and acknowledge support from grant ANPCYT-PICT 0478/14. GJML also acknowledges support from grants PIP-CONICET/2011 \#D4598. JLS acknowledge support from NASA grants NNX15AF19G and NNX17AC45G.

\end{acknowledgements}

% \vspace{5mm}
% \facilities{{\it Swift}(BAT, XRT and UVOT), XMM-{\it Newton}}



\bibliographystyle{aa}    %% bibliography style file aa.bst from A&A
\bibliography{/media/SEAGATE/listaref_MASTER}

\end{document}